# Searching for outliers in the *Chandra* Source Catalog

Dustin K. Swarm ,[1]⋆ C. T. DeRoo ,[1] Y. Liu[2] and S. Watkins[1]

[1]*Department of Physics and Astronomy, University of Iowa, 203 Van Allen Hall, Iowa City, IA 52242-1479, USA*
[2]*Iowa Initiative for Artificial Intelligence, University of Iowa, 103 South Capitol Street, Iowa City, IA 52242, USA*

## ABSTRACT

Astronomers are increasingly faced with a deluge of information, and finding worthwhile targets of study in the sea of data can be difficult. Outlier identification studies are a method that can be used to focus investigations by presenting a smaller set of sources that could prove interesting because they do not follow the trends of the underlying population. We apply a principal component analysis (PCA) and an unsupervised random forest algorithm (uRF) to sources from the *Chandra* Source Catalog v.2 (CSC2). We present 119 high-significance sources that appear in all repeated applications of our outlier identification algorithm (OIA). We analyse the characteristics of our outlier sources and cross-match them with the SIMBAD data base. Our outliers contain several sources that were previously identified as having unusual or interesting features by studies. This OIA leads to the identification of interesting targets that could motivate more detailed study.

## 1 INTRODUCTION

The pace of data acquisition in modern astronomy outpaces the ability of human astronomers to process using traditional methods. The *Nancy Grace Roman Space Telescope* and Vera C. Rubin Observatory alone are predicted to produce ∼500 petabytes of science data, which is 'several orders of magnitude more astronomical data than has been collected in human history' (National Academies of Sciences, Engineering and Medicine 2021).

The development and application of big-data techniques in the astronomical context is critical to the continued success of astronomical teams in extracting science information from a deluge of data. This is recognized by many in the astronomical community, as evidenced by the Astro2020 Decadal Survey and associated white papers. Barry et al. (2019) outline the necessity of monetary investment in big-data-ready astronomical methods as well as training and experience in data science skills for astronomers. Ntampaka et al. (2019) describe opportunities for big-data techniques to be incorporated in the analysis of large data sets, pipeline optimization, and archival data, while also acknowledging the need for improved cross-disciplinary collaboration to ensure the interpretability and rigour of discoveries driven by these techniques. The Astro2020 decadal report (National Academies of Sciences, Engineering and Medicine 2021) predicts that data science techniques 'could lead to transformative discoveries' in the coming decade and states that preparation for working with large data sets is necessary for astronomical progress.

One class of big-data methods that has an increasingly important impact on scientific discovery in astronomy is machine learning (ML).[1] ML is a class of computer algorithm that 'learns' to do a complex task through iterative training. ML techniques are broadly classified in two categories: supervised ML and unsupervised ML. In supervised ML, the algorithm is trained to do a task on data for which target variables are already known. This is akin to the algorithm checking its answers against a provided answer key and learning from its mistakes until it can perform the required task at some benchmark accuracy threshold. Unsupervised ML algorithms are trained using data for which the target variables are not known a priori, allowing the algorithm to find hidden patterns or structure in the data.

Common tasks for supervised ML algorithms include classification and regression. There are many examples of supervised ML algorithms being applied in astronomical contexts. The random forest classification algorithm for transients in the Palomar Transient Factory (Bloom et al. 2012) is an example of a supervised ML algorithm. Supervised ML algorithms have been used to classify and extract information from supernova light curves (e.g. Lochner et al. 2016; Charnock & Moss 2017; Narayan et al. 2018). Convolutional neural networks were used to predict the probability of complex Faraday spectra (Brown et al. 2019) and to classify variable star light curves in data from the Optical Gravitational Lensing Experiment (Szklenár et al. 2020).

Unsupervised ML algorithms are useful tools in tasks such as clustering or dimensionality reduction (Baron 2019). An unsupervised clustering algorithm has been used for feature learning in light curves of variable stars (Mackenzie, Pichara & Protopapas 2016). Self-organizing map algorithms were used to find unusual quasars (Meusinger et al. 2012), post-starbust galaxies in Sloan Digital Sky Survey (SDSS) data (Meusinger et al. 2017), and to uncover underlying temporal characteristics in variable *XMM–Newton* sources (Kovačević et al. 2022).

A variation of unsupervised clustering algorithms that can be of great use for astronomers is outlier identification algorithms (OIAs). An OIA learns the patterns in the data and searches for objects that break the patterns in some way. Astronomy has a heritage of new discoveries originating from outlier detection (e.g. Jocelyn Bell's discovering of pulsars). This makes OIAs not only a valuable set of astronomical tools for finding interesting research targets, but also a possible avenue for new discoveries.

---

⋆ E-mail: dustin-swarm@uiowa.edu
[1] See Baron (2019) for a detailed overview of various ML methods in the context of astronomy.

An example of an OIA applied in an astronomical setting is the work of Baron & Poznanski (2017) (hereafter BP17). BP17 adapt the unsupervised random forest (uRF) model introduced in Shi & Horvath (2006) to search for outlier spectra in the 12th data release of SDSS galaxy spectra. The SDSS catalogue is a large data set (∼2 million galaxies) that has been well characterized, including identification of unusual (i.e. outlier) galaxy spectra. This offered BP17 a useful benchmark to verify the performance of their outlier algorithm.

Their uRF-based outlier identification algorithm sorts input information (in this case, SDSS galaxy spectra) according to data values. Similar input is clustered together, while outlier input is isolated (see Section 4.1 for a more detailed explanation). BP17 applied their outlier identification algorithm to their sample of spectra and examined the 400 'weirdest' galaxy spectra as identified by the algorithm.

BP17 found that many outliers identified by the algorithm were indeed previously recognized outlier galaxy spectra, showing that the algorithm accurately identifies outliers. In addition to the known outliers, the algorithm also revealed several outlier spectra that had not been previously identified as unusual. BP17 classified the outlier spectra as examples of galaxies with unusual velocity structures, galaxies with unusual emission lines or emission line ratios, galaxies hosting supernovae, and spectra with instrumental errors. Their discovery of previously unidentified outliers illustrates the power of outlier identification algorithms in driving new research and opening avenues of potential discoveries.

Following the methodology of BP17, we search for interesting X-ray sources in the *Chandra* Source Catalog v2.0 (CSC2), a data base of X-ray objects observed by the *Chandra X-ray Observatory* (*CXO*) and formulate a publically available catalogue of unusual X-ray sources in the CSC2. Here we report on our applications of ML algorithms to the *Chandra* Source Catalog and discuss relevant issues facing astronomers interested in adopting ML in their research. In Section 2 we describe the *Chandra* Source Catalog and our analysis data. In Section 3 we detail our principal component analysis (PCA) of our selected CSC2 sources. In Section 4.1 we describe our application of the BP17 uRF to our CSC2 sources. In Section 5 we present our efforts to characterize the outlier sources found using the uRF.

## 2 DATA

While the *CXO* is not a survey mission, its two decades of operation have yielded hundreds of thousands of targeted and serendipitous detections. The CSC2 contains data for over 300 000 X-ray point-like and extended sources from 10 382 targeted observations made from 1999-2014 using the ACIS and HRC-I instruments (Evans et al. 2020). Each source in the CSC2 has information associated with three data tables: Per-Observation Detections Table, Stacked Observation Detections Table, and Master Sources Table.[2]

With the CSC2, point-like and extended sources were identified in 'detect stacks:' co-aligned, co-added observations taken with the same instrument and with pointing direction within 1 arcmin, allowing for the detection of fainter sources than the CSC v.1. Each 'master source' in the CSC2 has a single entry in the Master Sources Table and one or more entries in the Per-Observation Detections and Stacked Observation Detections tables corresponding to source detections in each observation and detect stack.

The Per-Observation Detections Table contains measured values for detection properties extracted from each observation in which the master source is detected. The Stacked Observation Detections Table contains measured values for detection properties extracted from the co-added observation stacks. The Master Sources Table contains measured and estimated source properties that represent 'best estimates' from a Bayesian block of observations with similar flux properties. The Master Sources Table alone contains >500 measured and estimated X-ray source properties.[3] CSC2 table data can take the form of strings, numbers, or Boolean values (see Table 2 for example CSC2 data). Each master source also has 40 associated Level 3 data products (e.g. spectra, event files, and images) (Evans et al. 2010, 2020). The information contained in the three data tables along with the Level 3 data products total nearly 36 TB of data in the CSC2.

### 2.1 Data selection and cleaning

In this analysis, we consider CSC2 master sources with flux significance[4] (significance in the Master Sources Table) ≥7.5. These are sources with high flux significance and are thus more likely to be sources for which (1) master source properties are well-constrained and (2) quality data facilitating follow-up analysis are already available. We conduct analyses on 27 (of >500) independent Master Sources Table properties. We selected these 27 to represent relevant summary properties related to the physical sources in the categories of location, significance, instrument (ACIS) usage data, colour, and parameters derived from fits with spectral models[5] (specifically power law, blackbody, and bremsstrahlung).

A difference between the SDSS spectra used by BP17 and our CSC2 data is the sparsity of much of the CSC2 Master Source table values. Missing information in our CSC2 data generally fell into one of three categories: hardness ratio, instrument observation time, or best-fitting spectral model information. The hardness ratio HR between two energy bands $h$ and $l$ is defined as:

$$HR_{h,l} = \frac{h - l}{h + l}, \quad (1)$$

where $h$ is the number of counts in the high energy band and $l$ is the number of counts in the low energy band. The CSC provides hardness ratios for three ACIS energy bands: soft (0.5−1.2 keV), medium (1.2–2.0 keV), and hard (2.0–7.0 keV), and hence provides three hardness ratios (hard_hs, hard_hm, hard_ms). Missing hardness ratios in the CSC2 data indicate that energy filters were used in observations in which the source was detected (CXC, 2021). Missing instrument observation time indicates that the source was not observed with that instrument. Missing values in spectral model parameters indicates that the source had <150 total summed background-subtracted photon counts in the 0.5–7.0 keV band (CXC, 2019), making model parameters ill-constrained by spectral fitting.

---

[2]See https://cxc.harvard.edu/csc2/organization.html for a detailed description of the source detection process and contents of each data table.

[3]Master Sources Table properties are alphabetically listed and described at https://cxc.harvard.edu/csc/columns/master_alpha.html

[4]The flux significance is an estimate of the ratio of the flux measurement to the average error in flux measurement (see https://cxc.harvard.edu/csc/columns/significance.html#flux_sig)

[5]See https://cxc.harvard.edu/csc/columns/spectral_properties.html#specmodfits for a detailed description of the spectral model parameters.

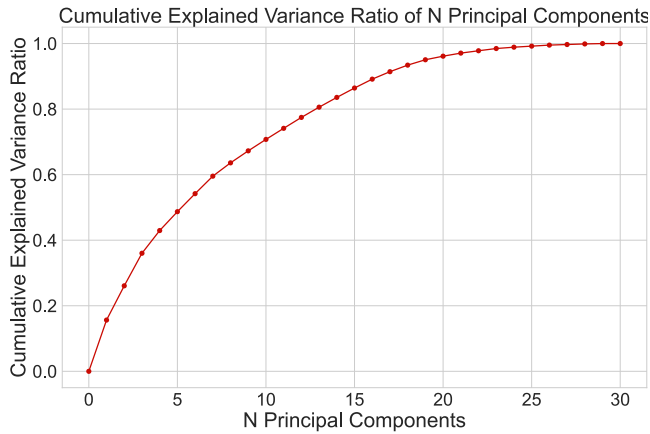

**Figure 1.** The cumulative explained variance ratio for principal components output by a PCA applied to the cleaned and scaled CSC2 sources. The cleaned CSC2 data used in this analysis consists of 30 features, and thus the full PCA produces 30 principal components.

We conduct preliminary data cleaning, removing sources with hardness ratio values that are null or unconstrained (i.e. at the bounds of ±1). After this preliminary cleaning we are left with an analysis data set of ∼ 35 000 sources.

Among these 35 000 sources, ∼ 50 per cent lack values for features related to automatic model fits (power law, blackbody, and bremsstrahlung). This makes the commonly used cleaning technique of row-deletion (i.e. removing sources with a null value in any feature, as in our preliminary cleaning of sources missing hardness ratios) undesirable for cleaning missing model-fit information.

Another common cleaning practice is imputation, a method in which null values are replaced by some other user-chosen value. Common choices for imputation values are 0, the feature-by-feature mean value of the non-null data, or a value determined using a regression method. However, both imputation and row deletion eliminate information that is implicit in the missingness of the data: the fact that the sources are missing data for a given feature could be important in our outlier analysis.

In order to preserve the information implicit in the missingness of the data while still formatting the data in a way that is usable by SCIKIT-LEARN analysers (Pedregosa et al. 2011), we employ a cleaning method known as 'dummy-variable imputation' or 'dummy-variable adjustment'. Dummy-variable imputation is a method in which a new variable is created to track which sources are missing values prior to imputation. Our dummy variables, named in the convention `missing_[variable name]`, took on the values of 1 if the source had a null value for that variable and 0 if the source had a non-null variable value. After the dummy variable is made, we impute a new value for the nulls. Because imputing a mean value or regression-predicted value would mask outliers in our data, we chose 0 as our imputation value.

This cleaning method resulted in three dummy variables: `missing_models`, indicating a source had no information in features relating to the automatically applied emission models; `missing_err_ellipse_ang` for eight sources without information of the offset angle for the source position error ellipse; and `missing_acis_time` for sources that were not observed with the ACIS instrument. The addition of these three dummy variables bring the total number of data features analysed in our work to 30 features per source (in Section 4.3 we discuss our finding that the dummy variables do not bias the results). A listing of the 30 analysis features can be seen in Fig. 5.

## 3 PRINCIPAL COMPONENT ANALYSIS OF THE CSC

We performed a principal component analysis (PCA) of our cleaned CSC2 sources. A PCA is a commonly used dimensionality reduction method that performs a singular value decomposition to transform data from one basis (the input features) to an orthonormal basis in the directions of maximal variance of the input data (Baron 2019). The first principal component (PC) lies along the axis of greatest variance in the data. The $i$th PC is then calculated to be orthogonal to the $i − 1$ previous PCs and in the direction that maximizes the remaining variance of the data. While $n$-dimensional data has $n$ PCs, it is common practice to use the first $m$ PCs ($m < n$) such that the total fraction of variance explained by the $m$ components satisfies a user-defined criterion (e.g. explaining 50 per cent of the variance). In this way the PCs allow users to reduce the dimensionality of their data.

### 3.1 Data scaling

Because the PCs are assigned to maximize the variance, features with orders-of-magnitude difference in variance will heavily influence the PCA. For example, the `acis_time` and `likelihood` features have standard deviations that are orders of magnitude larger than, for example, the hardness ratios (`hard_hs`, `hard_hm`, `hard_ms`). A PCA on the CSC2 data left in this state results in a first PC lying almost entirely along the `acis_time` feature and explaining >98 per cent of the variance in the data. Clearly the CSC2 sources cannot simply be characterized by the amount of ACIS livetime for their observations.

To overcome this disparity in variance we scale the data prior to application of the PCA. In this work, we use the `StandardScaler` object in the SCIKIT-LEARN library. The `StandardScaler` recentres and rescales each input variable $x$ to $z$ by

$$z = \frac{(x - \mu)}{\sigma}. \quad (2)$$

This results in data with $\mu_z = 0$ and $\sigma_z = 1$ for each input variable $x$, eliminating the bias seen in a PCA applied to the raw data.

### 3.2 CSC2 principal components

The full PCA of the scaled CSC2 sources outputs PCs that are frequently oriented along a blend of input variables rather than lying in the direction of a single variable. The cumulative explained variance of the PCs is shown in Fig. 1. The first five PCs (explaining 48.7 per cent of the variance) are illustrative in understanding the features that contribute to variance in the data, thus explaining the differences in sources. These PCs are summarized in Table 1. The first principal component is an axis describing the quality of the automatic model fits. The second principal component describes the general shape of the source spectrum through spectral hardness. The third PC describes the source's integrated model-specific flux, very generally describing how bright the source is. The fourth PC describes the uncertainty in the source's position measurement. The fifth PC is associated with model amplitudes, another general description of the brightness of the source. Thus, we interpret these PCs as describing the CSC2 data by how well simple models explain the observations, how hard the sources are, how bright the sources are, and how confident we are in their position. We plot the CSC2 data transformed to the PC space along the first five PCs in Fig. 2.

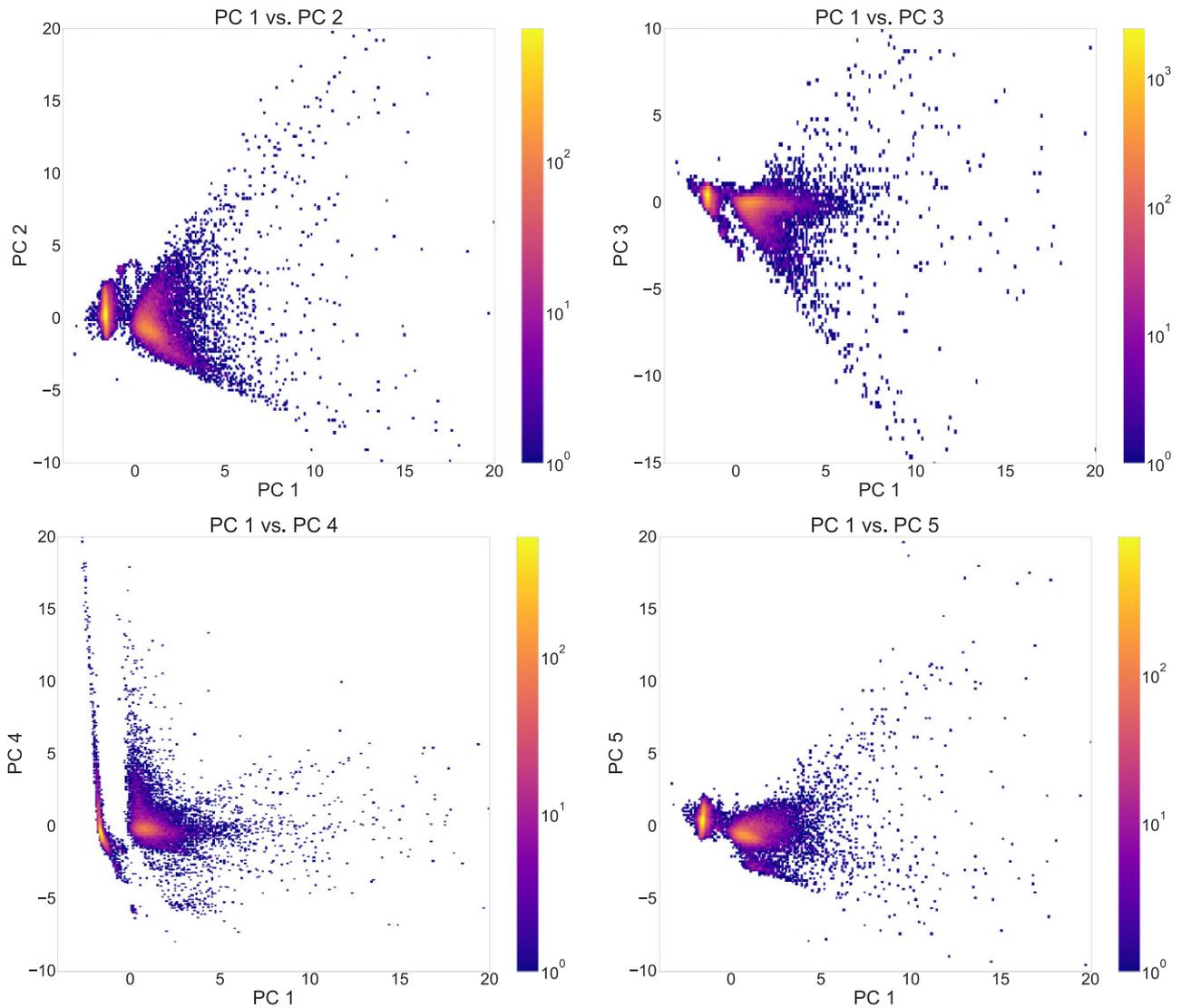

**Figure 2.** The CSC2 data transformed into the principal component basis and plotted in the first five principal components. Descriptions of the major feature constituents for these five PCs are given in Table 1. The bi-modality of the source distribution along PC 1 is based on the missing-models dummy variable; sources in the cluster centred in the negative region of PC 1 are missing model information, whereas sources in the cluster centred in the positive region of PC 1 contain model information.

## 4 OUTLIER IDENTIFICATION IN THE CSC2

This work employs an unsupervised random forest (uRF) as an outlier detection algorithm. Our analysis is implemented in PYTHON, leveraging the BP17 uRF tools[6] wrapped to be modular and compatible with numerical CSC2 data. The PYTHON files used in this analysis (along with our outlier catalogue) is publicly available on Github.[7]

### 4.1 Unsupervised random forests

An RF is a supervised ML technique first introduced by Breiman (2001). While RFs are useful in both classification and regression tasks, the uRF used in this work functions as a classifier.

RFs consist of a collection of decision trees, an algorithm introduced by Breiman et al. (1984) that sorts input data into labelled classes based on a series of binary cuts at threshold feature values of the form $feature \leq value$ that minimize the mixture in the resulting (i.e. child) branches (Loh 2011). This can be done by sampling all the data features, but more commonly this is achieved by sampling a randomized subset of the features (usually of size $\sqrt{n_{features}}$) (Shi & Horvath 2006; Nordhausen 2009). Cuts continue to be made on subsequent branches until the child branches consist of a single class or an early-stopping criteria is satisfied, with terminal nodes referred to as a 'leaf' (Loh 2011). The cuts used by the decision tree are the learned model for classifying the data.

Individual decision trees tend to overfit data; a decision tree can perfectly categorize the data in its training set but performs poorly in categorising new input data after training (Baron 2019). RF algorithms mitigate this overfitting problem by allowing many independently trained decision trees to vote on class assignments (Breiman 2001). A bootstrapped subsample is drawn for each tree from the training set, and the trees are grown in the manner described above. This results in a 'forest' of trees that filter input data based on unique random subsets of the training data and data features. The RF outputs the classifications with a plurality of votes from the constituent decision trees (Breiman 2001).

---
[6] https://github.com/dalya/WeirdestGalaxies
[7] https://github.com/dkswarm/CSCOutliers

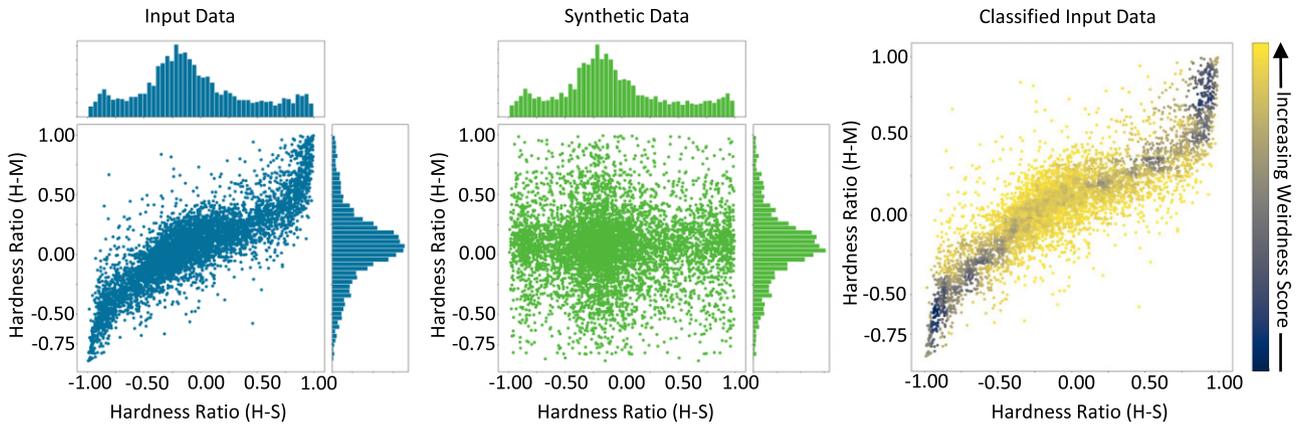

**Figure 3.** A comparison of real and synthetic data used in the uRF algorithm. Left-hand panel: Real data are input to the algorithm. Middle: Synthetic data are created to match the feature distribution of the real data (as seen in the histograms on the edge of each plot), but the feature covariance of the real data is ignored. Right-hand panel: The RF is trained to differentiate real data from synthetic data. The trained RF is then reapplied to the real data and weirdness scores are calculated. Weirdness scores are not unique and carry no intrinsic information about sources. Thus, the numeric values for the weirdness scores in this example plot are linearly scaled but not reported. Sources with high weirdness score (yellow) are those for which the RF is required to build in more rigorous cut criteria in order to differentiate real sources from synthetic sources.

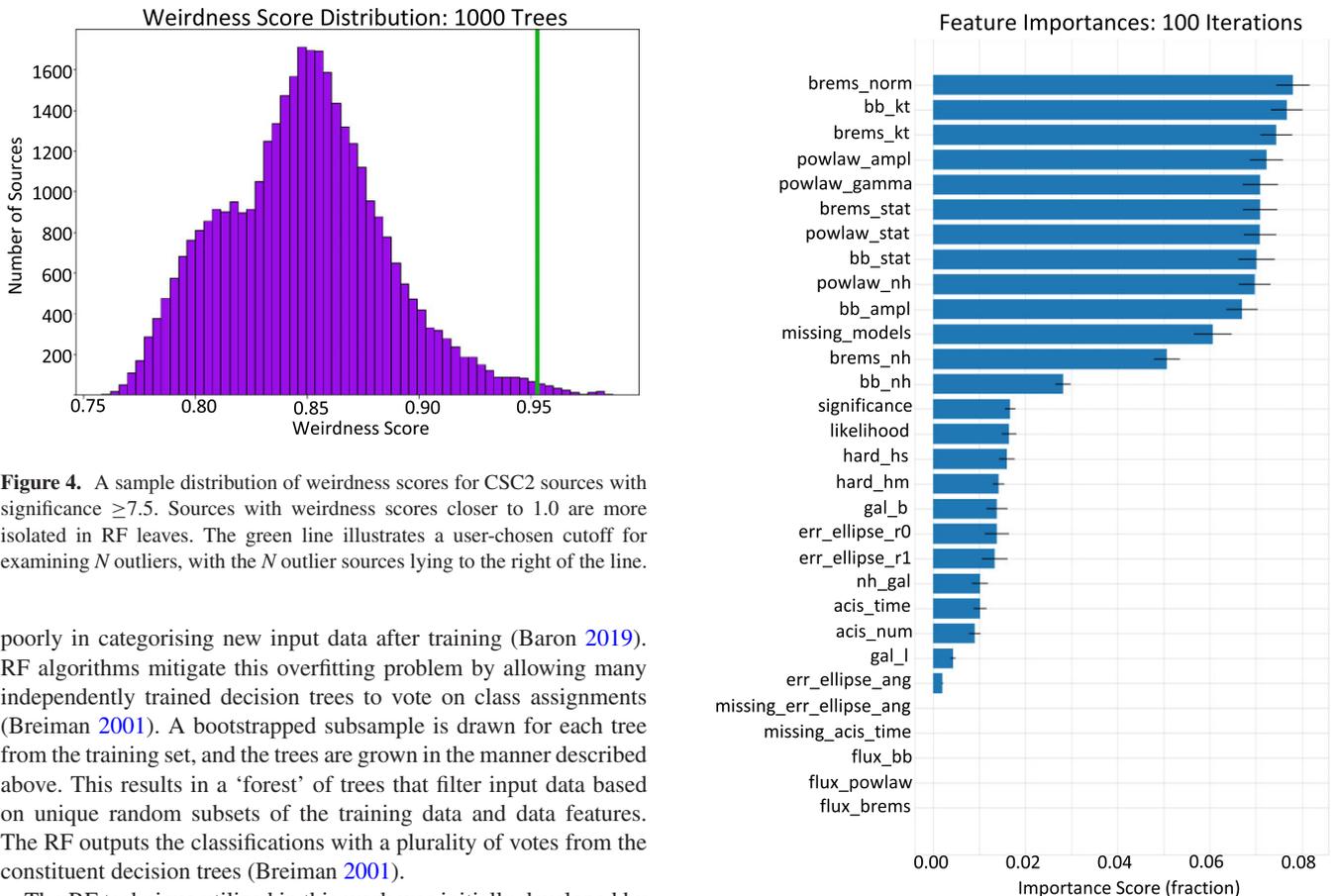

**Figure 4.** A sample distribution of weirdness scores for CSC2 sources with significance $\geq 7.5$. Sources with weirdness scores closer to 1.0 are more isolated in RF leaves. The green line illustrates a user-chosen cutoff for examining $N$ outliers, with the $N$ outlier sources lying to the right of the line.

**Figure 5.** The average feature importance for 100 RFs trained on data cleaned using dummy variable adjustment. Error bars indicate the $1\sigma$ standard deviation across the distribution of 100 iterations.

Th uRF technique utilized in this work was initially developed by Shi & Horvath (2006). In the uRF approach, the RF is trained to differentiate between real and synthetic data. The synthetic data are constructed to mimic the distribution of values for each feature seen in the real input data. However, the synthetic data do not replicate feature covariance that may exist in the real data (see Fig. 3 for an example of the relationships between real and synthetic data). After construction of a fake data set, the two sets are labelled ('real' or 'synthetic') and combined as a training set. The RF then builds a model to differentiate between real and synthetic sources, resulting in an RF that is sensitive to feature covariance.

**Table 1.** The variables with a weight ≥ 10 per cent for the first five principal components of the scaled CSC2 data. These five PCs explain ∼ 50 per cent of the variance in the CSC2 data.

| PC | Description | Primary CSC2 features in PC | Variable versus PC correlation |
|---|---|---|---|
| 1 | Quality of model fits | `bb_stat` | + |
|   |   | `brems_stat` | + |
|   |   | `powlaw_stat` | + |
|   |   | `missing_models` | − |
| 2 | Spectral hardness | `hard_hs` | + |
|   |   | `hard_hm` | + |
|   |   | `brems_nh` | + |
| 3 | Integrated energy flux | `flux_bb` | + |
|   |   | `flux_brems` | + |
|   |   | `flux_powlaw` | + |
| 4 | Source-position uncertainty | `err_ellipse_r0` | + |
|   |   | `err_ellipse_r1` | + |
| 5 | Model amplitudes | `bb_ampl` | + |
|   |   | `brems_norm` | + |

The trained RF can then be reapplied to the real data as an unsupervised clustering tool by examining the source composition of RF leaves. Sources that end in the same leaf take the same path through the decision tree, meaning they have similar observed properties and are well described by the same RF model.

BP17 used this principle to define a similarity measure for all source pairs in the RF. The similarity $S_{ij}$ between the *i*th and *j*th objects is the number of trees in which the pair share a leaf normalized by the number of trees in which the pair are classified as real sources

$$S_{ij} = \frac{N_{\text{leaf shared}}}{N_{\text{trees real}}}. \quad (3)$$

In this way, the similarity measure is the fraction of trees in which the *i*th and *j*th objects end in the same leaf. The 'distance' between any two objects is then

$$D_{ij} = 1 - S_{ij}. \quad (4)$$

The $D_{ij} \in [0, 1]$, with $D_{ij} = 0$ for sources that always land in the same leaf and $D_{ij} = 1$ for objects that never end in the same leaf. By definition $D_{ii} = 0$.

Finally, BP17 measure an object's uniqueness, what they dub a 'weirdness score,' as the average distance $D_{ij}$ from the *i*th object to all other objects:

$$W_i = \sum_j \frac{D_{ij}}{N_{\text{objects}}}. \quad (5)$$

The weirdness scores also take on the values $W_i \in [0, 1]$. The weirdness score contains no innate information about a source and is non-unique; RFs created with different initial states (i.e. seed states) will output different numerical values for the weirdness score of the same object.

Sources that are similar to many other sources will have a low average distance to the rest of the population and a low weirdness score; conversely, objects that are largely dissimilar from other sources will have a higher average distance to the rest of the population and a weirdness score closer to 1. Thus, outlier sources are those at the right tail of a weirdness-score distribution (see Fig. 4).

### 4.2 Implementation and hyperparameter selection

In initial testing of our adapted algorithm with the CSC2 data, we found that the process was memory intensive, which limited our analysis to <7000 sources at a time without inducing memory errors in a standard laptop computer. This necessitated the creation of a process for applying the algorithm to the entire CSC2 data set (∼ 35 000 sources). Our solution was to process the sources in smaller chunks. To begin, $N_{\text{train}}$ sources (where $N_{\text{train}}$ is the number of real sources in the training set) are randomly selected from the data set and used to construct the training set. After the RF is trained, the training sources are returned to the rest of the sources. At that point, the data set is divided into chunks of $N_{\text{train}}$ sources, and the trained RF is applied to each chunk for classification. After classification, weirdness scores are calculated as described in Section 4.1. A limitation of this method is that the weirdness score for an object in one application of the algorithm is really a measure of the source's similarity to the other sources in its chunk rather than the source's similarity to the rest of the data set. To address this issue and the inherent non-uniqueness of the weirdness score (see Section 4.1), we reapplied the chunking method *N* times using *N* different seed states and tracked the weirdness scores in each application.

A key factor in the performance of any ML algorithm is the choice of algorithm settings, or 'hyperparameters.' Hyperparameters vary from algorithm to algorithm, but some examples of RF-specific hyperparameters are the method for calculating a branch's impurity and the number of decision trees in the forest. We optimised hyperparameter values for: (1) training set size $N_{\text{train}}$, (2) impurity calculation method (i.e. the method by which the RF will calculate the mixture of nodes when making cuts as discussed in Section 4.1) (3) the number of decision trees to be used in each RF, and (4) the minimum allowable number of sources in a leaf (i.e. a terminal node).

A standard metric for evaluating hyperparameters is measuring the accuracy of the algorithm for combinations of hyperparameter values, with > 80 per cent accuracy being a common threshold for selecting 'good' hyperparameter values. However, all hyperparameter combinations resulted in > 99 per cent classification accuracy for our algorithm, indicating that the algorithm's accuracy was not sensitive to these hyperparameters in our data. We instead chose the values of hyperparameters that maximized the repeatability of the outlier algorithm across 10 applications. In each application we recorded the top 200 outliers. We measured repeatability by counting the number of unique sources in the 10 applications and the number of sources that appeared in all 10 applications. An algorithm with high repeatability would result in 200 unique sources across the 10 applications and 200 sources that appear in all 10 applications. Conversely, an algorithm with low repeatability would result in 2000 unique sources across the 10 applications, and 0 sources that appear in all 10 applications. We chose hyperparameter settings to minimize the number of unique sources that appear in the top 200 across 10 applications and maximize the number of sources that appear in the top 200 in all applications.

The hyperparameter settings that resulted in the highest degree of repeatability were: (1) a training set size of 6000 sources, (2) Gini impurity as the impurity measure,[8] (3) 1000 decision trees in each RF, and (4) a minimum of 1 source per leaf (i.e. branches are allowed to split until the terminal nodes are homogeneous).

---

[8]See https://scikit-learn.org/stable/modules/tree.html#classification-criteria for a list of the available classification criteria for SCIKIT-LEARN decision tree classifiers

### 4.3 Bias study

We investigated the effect of the data cleaning method described in Section 2.1 on the performance of the outlier identification algorithm by examining the feature-importance rankings and classification bias of trained RFs.

*4.3.1 Feature importance rankings*

One benefit of RF algorithms is that it is easy to examine the importance each data feature plays in the RF's classification process. Features that frequently reduce the impurity of branches are given a higher importance (or weighting) in the RF's decision making, whereas features that do not greatly reduce the impurity of branches are given a lower importance.

In order to determine the overall importance the dummy variables played in RF classifications, we trained 100 RFs on data cleaned using the dummy variable adjustment. We averaged the feature importances for all 100 RFs and plotted the rankings with $1\sigma$ error bars in Fig. 5. This showed three groupings of data features: a group of model-specific features that generally had high importance, a middle group of features that had a much lower feature importance, and five features with a feature importance of $f = 0$. The only dummy variable with a non-zero feature importance was the missing-model variable, with an averaged feature importance of $f \sim 0.06$. This places the missing-model variable in the bottom of the most important feature population. We interpret this as suggesting that the missing-model dummy variable is a helpful marker in distinguishing between low-count sources with noisy feature values and synthetic sources with randomly generated feature values.

*4.3.2 Classification bias*

The second property we analysed was the classification bias of the algorithm. As stated in Section 4.2, the RFs exhibited a high degree of accuracy in classifying sources as 'real' or 'synthetic' for all hyperparameter combinations for both cleaned and uncleaned data. However, we wanted to determine if cleaned data introduced a classification bias toward real or synthetic sources.

To quantify the impact of our chosen cleaning method (see Section 2.1) on the classification bias of the outlier algorithm, we compared the performance of RFs trained on raw data (i.e. data with `null` values) to the performance of RFs trained on data cleaned using the dummy variable adjustment method. For both the cleaned and raw data sets, 100 RFs learned training sets comprised of an even mixture of real and synthetic data. The trained RFs were then applied to new data sets, again comprised of even mixtures of real and synthetic data. The trained RFs predicted the class for all sources in the training set, and the predicted classes were recorded. The classification bias $b_{\text{class}}$ for each RF is calculated as:

$$b_{\text{class}} = f_{\text{predicted real}} - f_{\text{known real}} = \frac{N_{\text{sources labelled real}}}{N_{\text{sources}}} - 0.5, \quad (6)$$

where $f_{\text{predicted real}}$ is the fraction of sources that the RFs predict to belong to the class of real sources, and $f_{\text{known real}}$ is the fraction of sources that actually belong to the class of real sources. The $f_{\text{predicted real}}$ is simply the number of sources labelled 'real' by the RF divided by the total number of sources. Because the training set is created to contain equal numbers of real and synthetic sources, $f_{\text{known real}} = 0.5$. We interpret a $f_{\text{predicted real}} = 0.5$, combined with the high degree of classification accuracy, to indicate no significant classification bias.

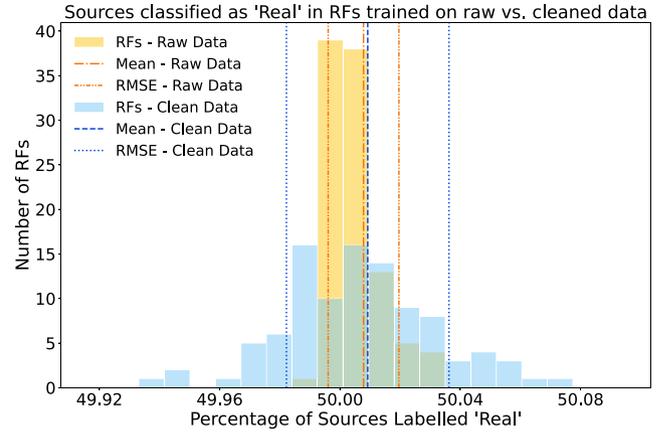

**Figure 6.** The distribution of output RF predictions in the classification-bias study. One hundred RFs were trained on an even mixture of real and synthetic sources. The trained RFs were then applied to a new set of data, again with a 50 per cent–50 per cent mixture of real and synthetic data. The output predictions for the test set were tallied, and the fraction of sources that were labelled 'real' was recorded. This plot shows a histogram of the percentage of sources labelled 'real' for 100 RFs trained on data cleaned using the dummy variable adjustment method and 100 RFs trained on raw data.

Fig. 6 shows a histogram of the percentage of sources predicted to be real for both RFs trained on raw data and RFs trained on data cleaned using dummy variable adjustment, along with the mean and $1\sigma$ standard deviation. Both sets of RFs skewed slightly toward mislabelling sources as real, but the skew never exceeded 0.1 per cent.

Given that (1) the dummy variables were not the most important features and (2) the classification bias measured in RFs trained on cleaned data never exceed 0.1 per cent, we determined that the dummy variable adjustment data cleaning procedure would not adversely affect the outcomes of our outlier identification algorithm.

### 4.4 CSC outliers

For the finalized stage of the outlier detection process, we created 100 RFs. Each RF was trained using a unique training set of CSC2 sources selected at random and combined with an equal number of synthetic sources generated from the selected CSC2 sources (see Section 4.2). The decision trees in the RFs sampled a randomized selection of five features $\left(\sqrt{n_{\text{features}}} = \sqrt{30}\right)$ at each node and chose the feature and value interval that resulted in the greatest reduction of the child-branch impurity until the leaves contained a uniform class.

The 100 trained forests were applied to the entire analysis population, and we recorded the 200 weirdest outliers from each forest. Our choice of 200 outliers with the highest weirdness score was motivated by convenience; it was the upper limit of sources we were willing to cross-reference with the SIMBAD Astronomical Database and other literature (see Section 5.3) by hand.

Across 100 applications of the uRF algorithm, 368 different sources appeared among the top 200 outliers. Of these 368 sources, 119 appeared in the top 200 in all 100 applications of the uRF algorithm. A machine-readable file containing our 119 outlier sources and their CSC2 properties is available online as a digital supplement, with an abbreviated version shown in Table 2. Interestingly, all of the 119 outliers that appeared in every uRF application had model information available, indicating that model information is a powerful discriminator in distinguishing between sources.

**Table 2.** Here we present the CSC2 data for 5 of the 119 sources identified as outliers in 100 applications of the uRF outlier identification algorithm. All CSC2 names are appended with '2CXO' (2CXO J...). Subsequent features are left with their CSC2 Master Sources Table designation. Master Sources Table property descriptions are provided in the CSC documentation (see footnote in Section 2 for a link to the property descriptions). A machine-readable version of our catalogue, including best-matched SIMBAD sources and source types, is available online.

| Outlier No. | 1 | 2 | 3 | ... | 118 | 119 |
|---|---|---|---|---|---|---|
| CSC2 Name (2CX0) | J004258.6 + 411527 | J004617.7-420751 | J010212.0-215209 | ... | J234349.4-151704 | J235842.8-322605 |
| gal_l (degrees) | 121.222 | 306.608 | 148.47 | ... | 66.5167 | 4.86367 |
| gal_b (degrees) | −21.5861 | −74.9605 | −84.1997 | ... | −70.3255 | −77.4019 |
| err_ellipse_r0 (arcsec) | 0.72 | 0.71 | 1.02 | ... | 0.71 | 3.46 |
| err_ellipse_r1 (arcsec) | 0.71 | 0.71 | 0.86 | ... | 0.71 | 2.27 |
| err_ellipse_ang (degrees) | 84.2 | 100.4 | 81.4 | ... | 179.8 | 99.9 |
| significance | 24.42 | 12.92 | 9.79 | ... | 23.44 | 8.57 |
| hard_hm | −0.1711 | 0.9107 | 0.9694 | ... | 0.8944 | 0.446 |
| hard_hs | −0.3198 | 0.9194 | 0.7245 | ... | 0.5084 | 0.0712 |
| nh_gal ($N_{\text{Hatoms}} \times 10^{20}$ cm$^{-2}$) | 6.69 | 2.15 | 1.6 | ... | 1.96 | 1.17 |
| acis_num (observations) | 108 | 1 | 4 | ... | 3 | 1 |
| acis_time (ks) | 781.264 | 18.3819 | 193.701 | ... | 125.999 | 48.9372 |
| likelihood | 3987.61 | 1038.7 | 325.209 | ... | 2726.34 | 91.0819 |
| flux_powlaw (ergs s$^{-1}$ cm$^{-2}$) | 6.438e-18 | 2.622e-13 | 4.359e-14 | ... | 1.741e-13 | 2.522e-14 |
| powlaw_gamma | −2.03 | −1.103 | −1.924 | ... | −3.273 | 1.598 |
| powlaw_nh ($N_{\text{Hatoms}} \times 10^{20}$ cm$^{-2}$) | 2788 | 1.351e-05 | 2.136e-05 | ... | 4.797e-05 | 4.985e-08 |
| powlaw_ampl | 1.412e-08 | 1.227e-06 | 5.457e-08 | ... | 1.701e-08 | 4.369e-06 |
| powlaw_stat ($\chi^2/\nu$) | 16.078 | 1.249 | 1.5 | ... | 7.458 | 0.911 |
| flux_bb (ergs s$^{-1}$ cm$^{-2}$) | 4.879e-15 | 1.557e-13 | 3.195e-14 | ... | 1.022e-10 | 7.432e-15 |
| bb_kt (keV) | 0.8408 | 74.8 | 93.05 | ... | 39.37 | 0.3659 |
| bb_nh ($N_{\text{Hatoms}} \times 10^{20}$ cm$^{-2}$) | 9771 | 6.153 | 403.5 | ... | 2.15e-06 | 1.072e-07 |
| bb_ampl | 9.56e-05 | 1.957e-08 | 3.246e-09 | ... | 1.155e-08 | 6.415e-05 |
| bb_stat ($\chi^2/\nu$) | 16.077 | 1.302 | 1.77 | ... | 9.196 | 2.08 |
| flux_brems (ergs s$^{-1}$ cm$^{-2}$) | 1.379e-15 | 2.204e-13 | 4.102e-14 | ... | 1.264e-13 | 2.46e-14 |
| brems_kt (keV) | 2.434 | 100 | 100 | ... | 100 | 8.749 |
| brems_nh ($N_{\text{Hatoms}} \times 10^{20}$ cm$^{-2}$) | 8703 | 852.1 | 2108 | ... | 4434 | 9.131e-09 |
| brems_norm | 0.0002507 | 0.0001804 | 5.621e-05 | ... | 0.0003529 | 5.823e-06 |
| brems_stat ($\chi^2/\nu$) | 16.077 | 3.579 | 2.385 | ... | 8.953 | 1.013 |

## 5 OUTLIER POPULATION ANALYSIS

The choice of BP17 to use SDSS spectra in their uRF outlier analysis was partially motivated by the existence of catalogues of unusual SDSS galaxy spectra due to the extensively studied nature of the SDSS data releases. As mentioned in Section 1, the uRF algorithm was successful in identifying previously known outliers while also outputting outlier spectra that had not been recognized a priori.

We are unaware of a pre-existing catalogue of documented outliers in the CSC2. We therefore explore the characteristics of our outlier sources in three phases to investigate ways the outliers differ from the rest of the CSC2 analysis population. In Section 5.1 we compare the distribution of CSC2 feature values for outliers and non-outliers in our study population. In Section 5.2 we examine the location of outlier sources in the CSC2 data after a transformation to the PC space of Section 3.2. In Section 5.3 we match CSC2 outlier sources to counterparts in the SIMBAD data base (Wenger et al. 2000) and discuss SIMBAD sources that are identified as unusual by previous studies.

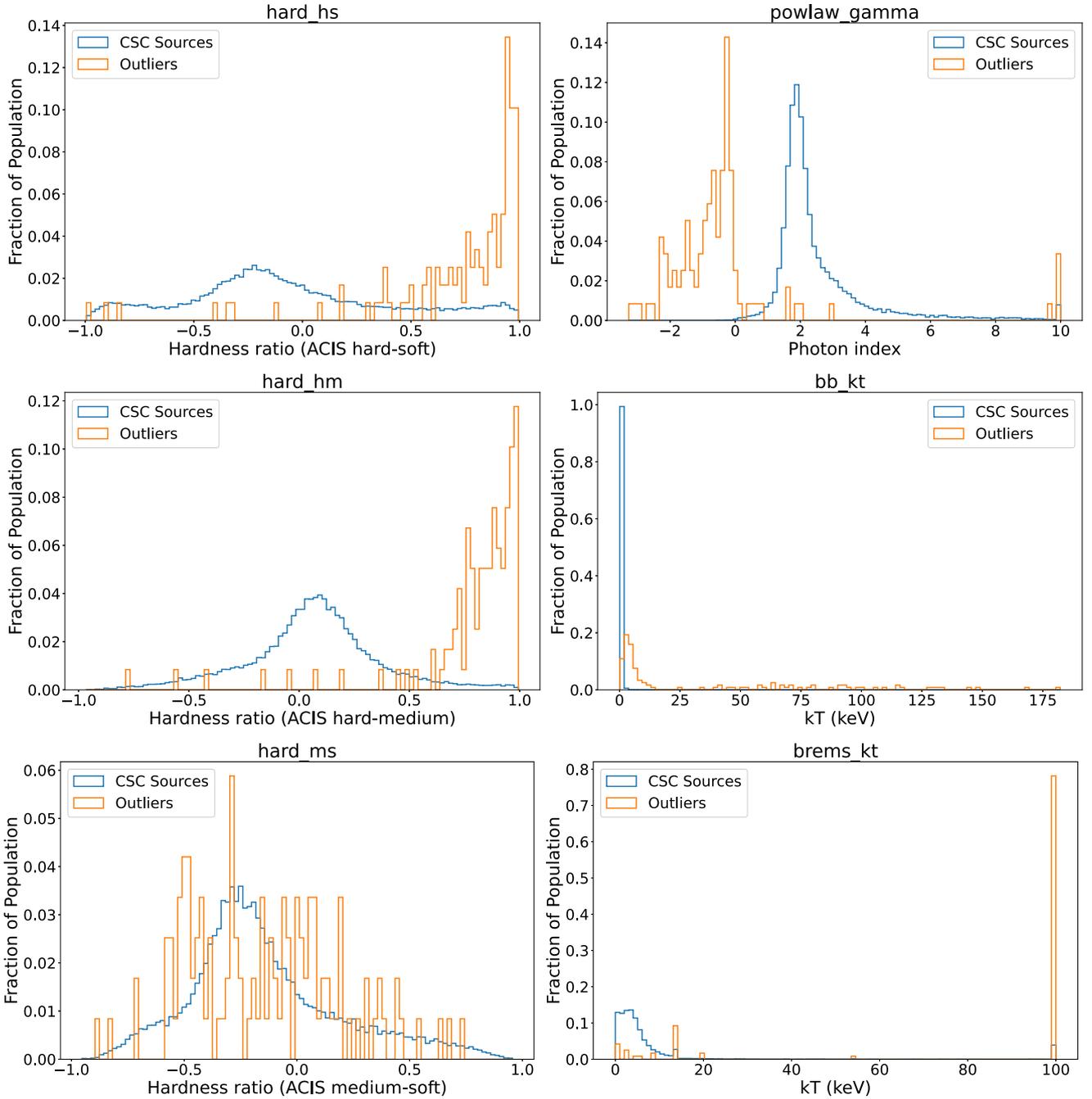

**Figure 7.** The feature distributions of the outlier sources (orange) plotted with those of the rest of the CSC2 sources analysed (blue) for features relating to spectral shape.

### 5.1 Outlier properties

We began by investigating the feature distribution of the outliers as a population distinct from the rest of our CSC2 data. To do this we plotted histograms of the outliers along with the non-outliers in each feature. The histograms were normalized to allow comparison of the distributions.

The most obvious difference between the outlier sources and the rest of the CSC2 sources is in their value distributions of features relating to the overall spectral shape. This comparison reveals that the outliers are predominately hard X-ray sources, as seen in the distributions of the $HR_{h-s}$, $HR_{h-m}$, power law photon index $\Gamma$ ($F_E \propto E^{-\Gamma}$), and blackbody and bremsstrahlung $kT$ (see Fig. 7).

Differences between most of the feature distributions of the outliers and non-outliers were not easily distinguishable by eye. For all feature distributions we conducted a two-sample (outliers and non-outliers) Kolmogorov–Smirnov test[9] and a two-sample

---

[9] Two-sample K–S test: https://docs.scipy.org/doc/scipy/reference/generated/scipy.stats.ks_2samp.html

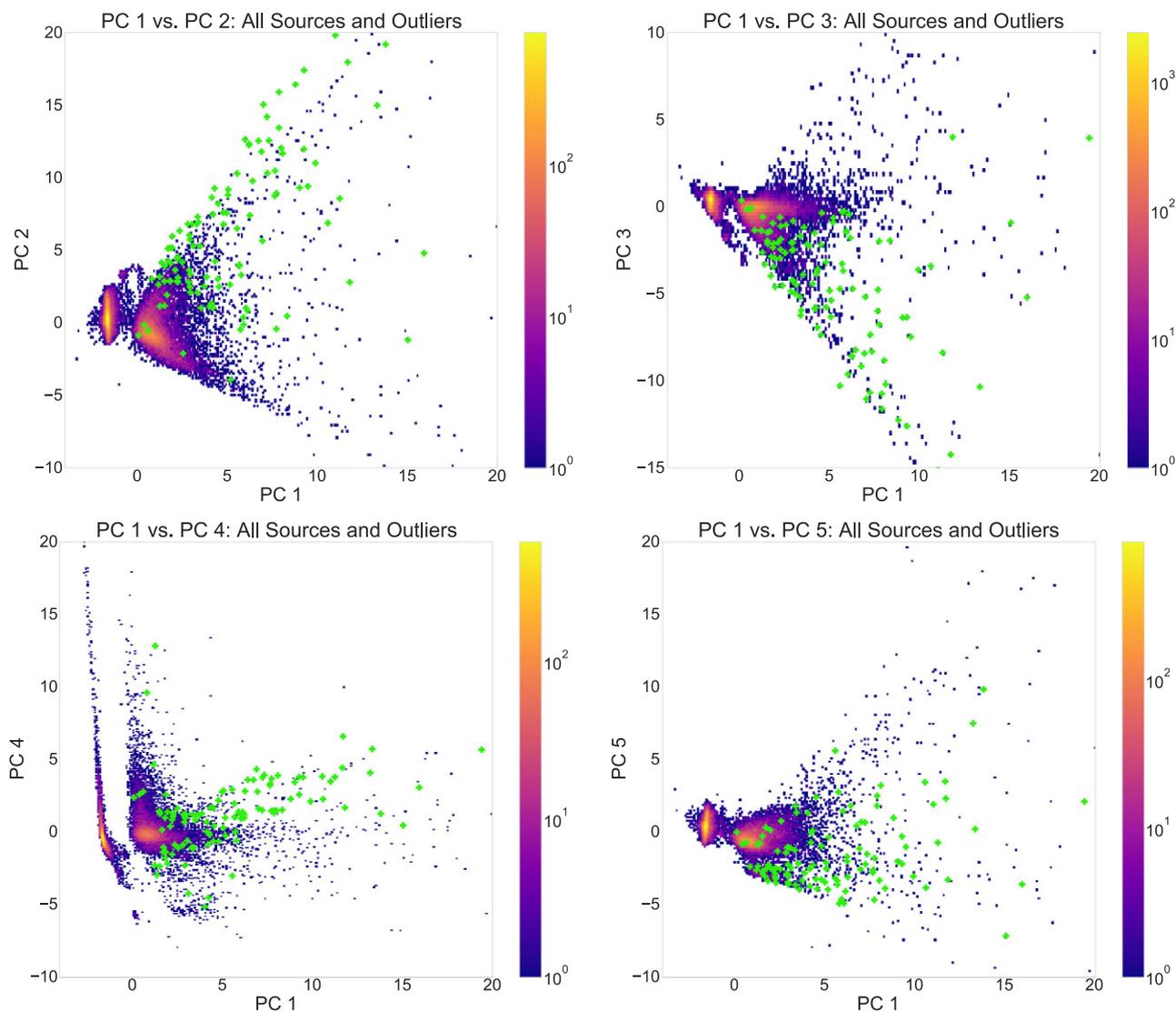

**Figure 8.** The CSC2 sources transformed to the PC space and plotted in the PC combinations shown in Fig. 2. The 119 outlier sources are plotted as green crosses.

Anderson–Darling test[10] (Hodges 1958; Scholz & Stephens 1987) using functions included in SCIPY. Both are tests of the null hypothesis – that two populations come from the same underlying distribution – and output a *p*-value used to determine the significance level $\alpha$ at which the null hypothesis can be rejected. More specifically, the *p*-value is the probability of obtaining an outcome at least as extreme as that observed if the null hypothesis is true. If $p \leq \alpha$, it is unlikely that the two populations are drawn from the same underlying distribution, and the results are considered statistically significant at significance level $\alpha$.

For the two-sample K–S test, the distributions of the outliers differed from the distributions of the CSC2 analysis data set for all features at $p < 0.001$, with the exception of err_ellipse_ang ($p = 0.291$) and hard_ms ($p = 0.060$). In the two-sample A–D test, the outlier distributions differed from the distributions of the CSC2 analysis data set for all features at $p < 0.001$ except gal_b ($p = 0.005$), err_ellipse_ang ($p = 0.094$), and hard_ms ($p = 0.070$). The K–S significance for err_ellipse_ang is too large to reject the null hypothesis, but the A–D significance for that feature would allow the null hypothesis to be rejected at $\alpha = 0.1$.

Applications of the two-sample K–S test and two-sample A–D to random draws of 119 sources from the full CSC2 data show the tests are sensitive to population similarities on the scale of our outlier catalogue. Both tests consistently identify the randomly drawn samples as having the same underlying distribution as the CSC2 analysis data set. We interpret this as indicating that, as a population, the outliers statistically differ from the rest of the CSC2 analysis data set in all features tested at $\alpha = 0.1$, with the exception of error ellipse angles in the K–S test. This result is not surprising given that these sources were chosen as outliers by the algorithm, but it does not offer deeper understanding into why our sources were chosen as outliers.

### 5.2 Outliers in the CSC2 PCA

Another analysis path for understanding the characteristics of the outlier sources was to look at where they fall in the PC space of

---

[10]*k*-sample A–D test: https://docs.scipy.org/doc/scipy/reference/generated/scipy.stats.anderson_ksamp.html

Section 3.2 in comparison to the non-outlier remnant of the analysis population. The PC space illustrated in Fig. 2 shows that the CSC2 sources as a whole do not separate into distinct groupings of sources when transformed to the PCs; there are clear trends in the groupings of the CSC2 sources, but the groups are not discontinuous.

When comparing the population distributions of the outliers and the non-outliers for individual PCs, we see the outlier sources do not follow the distributions of the non-outliers in the individual PC components. This is not surprising when considering the PC values are linear combinations of the CSC2 data feature values, which our statistical distribution tests showed outlier sources were selected from a separate underlying distribution for all data features.

In Fig. 8 we overplotted the location of our outlier sources in the PC-space projections for the same PC combinations examined in Section 3.2. The outliers are not isolated from the non-outliers by the PC transformation; there are non-outlier sources that appear in similar locations in the PC space. However, we note that the bulk of outlier sources are clustered together, rather than spread throughout the CSC2 distribution of the PC space. This strongly suggests the outlier sources have underlying property distributions that follow distinct trends.

The outliers generally have positive values in PCs 1 and 2 (representing model fit statistic and spectral hardness, respectively) and have negative values in PC 3 (representing model flux). The outliers are more evenly distributed in PCs 4 and 5 (source positional uncertainty and model amplitudes, respectively) with a skew to positive values in PC 4 and a skew to negative values in PC 5.

The positive values of the outliers in PC 1 indicates that the outlier sources have model fit $\chi^2$ statistics that are greater than the average for the entire CSC2 analysis population,[11] suggesting they are more poorly described by simple emission models. The positive values in PC 2 indicate the outliers have a harder-than-average spectrum, in line with the heavy skew toward hard sources seen in Section 5.1. The negative values in PCs 3 and 5 indicate the outliers are lower flux sources compared to the average for the CSC2 analysis population. The positive skew for the outliers in PC 4 suggests that there is a greater source positional uncertainty for the outlier sources than the average of the analysis population.

This analysis process again revealed general trends in the outliers without clearly differentiating a single set of data features as the defining characteristic of outlier sources. However, this analysis did illustrate that the outlier sources follow internal trends resulting in clustering within the PC space.

### 5.3 Catalogue cross-referencing

Because the CSC2 does not include source identifiers beyond CSC2 designations seen in Table 2, we utilized the SIMBAD data base (Wenger et al. 2000) to find source designations from other missions, object classifications, and bibliographic references. We cross-referenced the CSC2 coordinates for the outlier sources with source coordinates in the SIMBAD data base using a 3 arcsec SIMBAD coordinate search radius.

We manually verified the SIMBAD match for all catalogue sources that returned at least one SIMBAD match by searching through associated literature references to determine the SIMBAD object most likely to be correlated with our CSC2 catalogue object. Once a likely SIMBAD candidate was determined through literature review, we verified that the outlier source was the CSC2 object with the smallest angular separation from the SIMBAD object. When a SIMBAD object was confirmed to match a catalogue object, we logged the SIMBAD object classification. Fig. 9 shows the source types for our outlier sources based on the classifications of the matched SIMBAD objects.

---

[11]The PCA is applied to standardized data as discussed in Section 3.1.

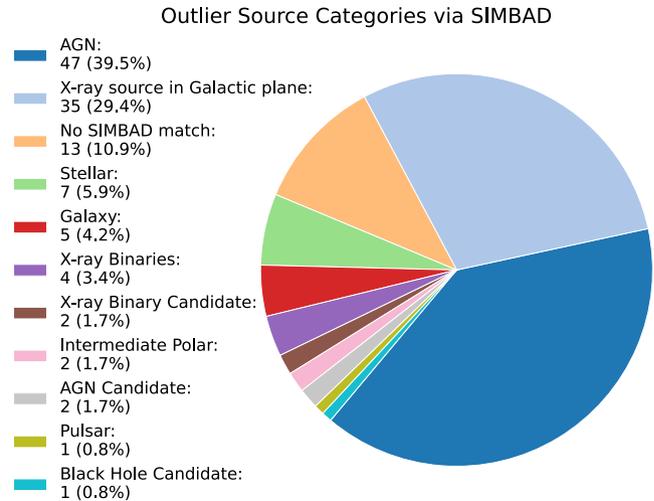

**Figure 9.** The source category composition of the 119 outliers found by the uRF algorithm. Object types were found by cross-referencing the source coordinates in CSC2 with objects in the SIMBAD data base.

Our outlier sources consist of a mixture of Galactic and extragalactic X-ray sources. AGN make up the largest object category in our catalogue (47 outlier sources), followed by X-ray point sources within the Galactic plane (35 outlier sources). There were 13 outlier sources with no matching SIMBAD object. As might be expected, the 13 outliers without a SIMBAD match trend toward the low end of the outliers' flux-significance distribution; 10 fall below the outliers' median flux significance.

The previously described investigations of the outlier CSC-feature distributions and the location of outliers in the PC space did not yield definitive indications into the differences between the outliers and non-outliers. However, our SIMBAD cross-matching did uncover several outlier objects that were identified as unusual by previous investigations, some of which were studied outside the X-ray waveband (e.g. radio and optical). Here we discuss a few of these previously identified interesting sources that also appear in our outlier catalogue.

#### 5.3.1 AX J1740.1-2847

AX J1740.1-2847 (Outlier 66 in Table 2) was first identified in 1998 as a faint, hard X-ray source in the *ASCA* Galactic center survey. Sakano et al. (2000) identified AX J1740.1-2847 as a slow X-ray pulsator. They observed slow pulsations with a 729 s period, which they state is consistent with the spin period of a white dwarf or neutron star. This, along with spectral fits, led them to suggest that AX J1740.1-2847 is binary with a white dwarf or neutron star (X-ray binary pulsar) primary, favoring the neutron star hypothesis.

Muno et al. (2004) presents the spectral and variability study of X-ray point sources observed in early deep *Chandra* observations of the Galactic centre region. Here AX J1740.1-2847 is included in a table of unusual sources due to the slow, high-amplitude periodic modulations seen AX J1740.1-2847 in the X-ray.

Kaur et al. (2010) use *Chandra* and *XMM–Newton* observations of five low-luminosity X-ray pulsators identified in *ASCA* Galactic

plane observations, along with archival NIR/optical observations, in order to gain a clearer picture into the nature of these sources. Kaur et al. (2010) agree that all of their sources are most likely to be accreting white dwarfs or neutron star binaries. However, the *XMM–Newton* observations of AX J1740.1-2847 revealed a strong 6.4 keV line which, coupled with the long spin period and the NIR magnitude of the companion star, led Kaur et al. (2010) to view AX J1740.1-2847 as most likely an intermediate polar (IP), a class of accreting white dwarf system (Warner 1995).

Following a similar research path, Britt et al. (2013) study five X-ray sources in the *Chandra* Galactic Bulge Survey with bright optical counterparts exhibiting strong emission lines. Given the X-ray periodicity, iron lines, hard X-ray emission, and optical properties, Britt et al. (2013) classified AX J1740.1-2847 as a 'definitive IP.' This makes AX J1740.1-2847 one of only 71 confirmed IPs in Koji Mukai's IP catalog,[12] compared to 3408 confirmed cataclysmic variables (accreting white dwarfs) listed in the Open Cataclysmic Variable Catalog.[13]

*5.3.2 CXOGBS J174954.5-294335*

Johnson et al. (2017) was the first to identify CXOGBS J174954.5-294335 (Outlier 108) as a deeply eclipsing IP, along with its optical counterpart the eclipsing binary system OGLE BLG501.24 34934 described in Udalski et al. (2012). CXOGBS J174954.5-294335 exhibits deep eclipses in both the optical and X-ray data, making it only the second IP identified with deep eclipses in multiple bands. Johnson et al. (2017) also notes that CXOGBS J174954.5-294335 exhibited several outbursts in the optical OGLE data. Hameury & Lasota (2017) classifies these optical outbursts as dwarf novae, a phenomenon that is rare in IPs.

*5.3.3 HD 152270*

Outlier 64 matched the SIMBAD source HD 152270 (also commonly called WR 79), a Wolf–Rayet star in a binary with an O-type companion. HD 152270 has nearly 300 associated publications in the SIMBAD data base.

HD 152270 has garnered extensive study due in the star's winds. Seggewiss (1974) first noted unusual variable emission lines in the wind spectrum. Gayley, Owocki & Cranmer (1997) included HD 152270 in an analysis of sudden radiative breaking in colliding winds from hot stars. Studies such as Luehrs (1997) and Hill et al. (2000) have sought to explain the unusual emission lines through modelling them as arising from colliding winds. Massa (2017) studied the way the HD 152270 wind interacts with its surrounding cluster, NGC 6231. Fullard et al. (2020) explains the intrinsic polarization in the wind spectra as arising from an asymmetric collision area.

*5.3.4 3C 285*

Outlier 48 matches SIMBAD source 2MASX J13211781+4235153, a nearby FR II galaxy also known as 3C 285. This is another well-studied object, with 260 associated publications in the SIMBAD data base.

3C 285 is notable as one of a small set of AGN observed to exhibit jet-induced star formation, or 'positive feedback.' A blue, star-forming region in its radio lobe was first reported by van Breugel & Dey (1993) and attributed to interactions between the radio jet and a nearby molecular cloud, with later studies by Salomé, Salomé & Combes (2015) on the star formation efficiency due to this positive feedback. Hardcastle et al. (2007) report on X-ray emission from the jet-induced star formation in 3C 285.

*5.3.5 IC 450*

IC 450 (Outlier 24) is a nearby Seyfert 1.5 AGN in a lenticular galaxy (Kharb et al. 2014), also commonly referred to as Mrk 6. IC 450 is currently referenced in more than 500 publications in the SIMBAD data base.

IC 450 is noted for having an unusual gas structure in the radio bubbles extending perpendicularly from the galaxy plane (e.g. Kharb et al. 2014) and complex X-ray absorption (e.g. Schurch, Griffiths & Warwick 2006). The perpendicular bubbles could be caused by precessing jets. Both Schurch et al. (2006) and Mingo et al. (2011) note complex structure in the gas observed in the X-ray. Observations of IC 450 using the VLBA show curved radio jets, which Kharb et al. (2014) interprets along with *Hubble* and X-ray observations to be evidence of jet precession.

IC 450 exhibits a high degree of variability in X-ray absorption and in polarization. The variable X-ray absorbing column is explained by Schurch et al. (2006) as a warm absorber rather than a partial covering model. However, Mingo et al. (2011) view the variability in X-ray absorption as due to clumpy accreting gas close to the AGN. IC 450 exhibits significant broad emission line variability that Afanasiev et al. (2014) attributes to ISM-polarization contributions to the observed polarization in IC 450.

*5.3.6 R Aqr*

Outlier 118 matches R Aquarii, the most studied symbiotic stellar system. R Aqr contains a Mira variable star primary, a white dwarf companion, and surrounding planetary nebula with equatorial structure and precessing bipolar jets (Bujarrabal et al. 2021). Observed in the radio, optical, UV, and X-ray (Kellogg et al. 2007), R Aqr is referenced in more than 750 publications in SIMBAD.

Because R Aqr is relatively close (∼200 pc), with features that evolve with time-scales on the order of years, R Aqr offers a unique laboratory for observing binary accretion, evolution, and jets (Kellogg et al. 2007). As of 2007, R Aqr is also one of only two white dwarfs that exhibit X-ray jets (Kellogg et al. 2007). R Aqr has even been used to study the gravitational effects of compact companions on stellar winds (e.g. Bujarrabal et al. 2018).

The highly collimated jets have long attracted attention. Paresce & Hack (1994) refers to R Aqr as 'a bizarre symbiotic' and notes that the highly collimated jets are common in young stellar objects but rarely seen in evolved systems. Potential explanations for the collimation include a stellar wind restricted by an ambient medium or a magnetic field (Burgarella, Vogel & Paresce 1992), but the exact source of the collimation has not been confirmed (Melnikov, Stute & Eislöffel 2018). Using *HST* to measure emission line ratios in the R Aqr jet, Melnikov et al. (2018) detected two different outflows, most likely with different mechanisms. Melnikov et al. (2018) also reported a 'wiggling' pattern in the emission lines of the inner jet, a feature also

---

[12]https://asd.gsfc.nasa.gov/Koji.Mukai/iphome/catalog/alpha.html
[13]https://depts.washington.edu/catvar/index.html

## 6 CONCLUSIONS

We performed a principal component analysis of high significance master sources from the CSC2 to identify the features with the largest contribution to the observed variation within our data set. The majority of the variance in our CSC2 sources was explained through principal components primarily related to the fit statistics of the simple emission models, the shape of the X-ray spectrum, and the source positional uncertainty. We then presented the results of applying an unsupervised random forest algorithm used for outlier identification to the CSC2. Our methodology is based on the uRF algorithm utilized in BP17, and includes an additional discussion on how to select appropriate hyperparameters and assess classification bias in an astronomical context. The resulting outlier search yielded 119 sources that appeared among the 200 'weirdest' CSC2 sources in 100 applications of the uRF OIA.

We next analysed the CSC2 feature distribution and location in the PC space of our outliers to determine the ways in which they differed from the larger CSC2 population. The feature distribution of the outliers most strongly differed from the rest of the CSC2 sources within the hardness ratios and best-fitting power-law photon index. Within the PC space the outliers remained grouped together but were not significantly isolated from the rest of the CSC2 population.

We cross-referenced the CSC2 source coordinates with the SIMBAD data base to investigate the source objects. Nearly 70 per cent of our outliers were matched to AGN or X-ray point sources located within the Galactic plane. Eleven per cent had no match in SIMBAD; most of these sources appear among the lower-significance outliers, as may be expected for sources that have not appeared in previous studies. We briefly describe six outliers matched to previously identified unusual objects, motivating that it may be an effective method in identifying sources worth additional study. However, the majority of our outliers appear in multiple *CXO* pointings with long ACIS livetime but never the less have few or no dedicated publications linked in SIMBAD, making them prime targets for follow-up. Further analysis of these sources, particularly a subset located in the dynamic and complex Galactic Centre region, is the intended subject of a future investigation.

A potential improvement to future applications of the uRF OIA would be the inclusion of CSC2 best-fitting model parameter confidence limits. As discussed in Section 2.1, CSC2 Master Sources are automatically fit with simple emission models. The pipeline makes no discerning choices on physically appropriate models for a given source, which can lead best-fitting parameter values that are non-physical but informative, never the less. While fit uncertainty was evaluated by our OIA through the model fit statistics, inclusion of the CSC2-provided confidence limits for all best-fitting model parameters (along with confidence limits for other CSC2 Master Source properties considered in this work) would provide a more complete picture for the algorithm.

More generally, it is important to recognize that, especially when using unsupervised ML approaches, there is rarely a single 'correct' result. Our catalogue of 119 outlier sources is certainly not the definitive catalogue of unusual sources in the CSC2. Altering decisions during the development of the algorithm process (e.g. number of outliers tracked in each application of the algorithm, number of data features taught to the algorithm, number of applications of the algorithm, looser flux significance requirements, data cleaning methods) would undoubtedly provide different outlier lists. In unsupervised OIAs it is necessary balance choices based on domain knowledge with an open approach that allows the unsupervised method to potentially reveal unknown trends.

The deluge of data from observatories like *JWST*, *eROSITA*, *NGRST*, and the VRO will require astronomers to add ML techniques to their analysis tool kit. However, Barry et al. (2019) note that ML techniques rapidly evolve and carry a steep learning curve for astronomers seeking to effectively incorporate ML in their research. In this paper, we detail the choices and processes astronomers new to ML will have to make. By discussing in detail our process for hyperparameter testing and selection, data cleaning in sparse data sets, and bias evaluation, we hope to provide an example for astronomers who seek to incorporate ML into their research.

## ACKNOWLEDGEMENTS

This work is supported by the University of Iowa College of Liberal Arts and Sciences. This project is also supported by The Iowa Initiative for Artificial Intelligence. This project/material is based upon work supported by the Iowa Space Grant Consortium under NASA Award No. NNX16AL88H.

The authors acknowledge Randall Smith for informative and thoughtful discussions at the beginning of this project. The authors also acknowledge the help of Jared Termini in matching SIMBAD sources.

This research has made use of data obtained from the Chandra Source Catalog, provided by the Chandra X-ray Center (CXC) as part of the Chandra Data Archive. This research has also made use of the SIMBAD database, operated at CDS, Strasbourg, France.

## DATA AVAILABILITY

The *Chandra* Source Catalog and SIMBAD database are publicly available. The a machine-readable version of the outlier list is available online. The outlier list and Python files are available at github.com/dkswarm/CSCOutliers.

## SUPPORTING INFORMATION

Supplementary data are available at *MNRAS* online.

**Table 2.** Here we present the CSC2 data for 5 of the 119 sources identified as outliers in 100 applications of the uRF outlier identification algorithm.

Please note: Oxford University Press is not responsible for the content or functionality of any supporting materials supplied by the authors. Any queries (other than missing material) should be directed to the corresponding author for the article.

This paper has been typeset from a TeX/LaTeX file prepared by the author.